\documentclass[3p,times,twocolumn]{elsarticle}

\usepackage{ecrc}

\volume{00}

\firstpage{1}

\journalname{Nuclear Physics B Proceedings Supplement}

\runauth{Rafael L.~Delgado, Antonio Dobado and Felipe J.~Llanes--Estrada}

\jid{nuphbp}

\jnltitlelogo{Nuclear Physics B Proceedings Supplement}

\usepackage{amssymb}

\usepackage[figuresright]{rotating}

\newcommand{\Imag}{\mathop{\mathrm{Im}}}

\begin{document}

\begin{frontmatter}



\dochead{}

\title{Strongly interacting $W_LW_L, Z_LZ_L$ and $hh$ from unitarized one-loop computations }


\author{Rafael L.~Delgado}
\author{Antonio Dobado}
\author{Felipe J.~Llanes--Estrada}
\address{Departamento de F\'isica Te\'orica I, Universidad Complutense de Madrid, 28040 Madrid, Spain}

\address{}

\begin{abstract}
Recently, a new boson $h$ has been discovered at the LHC which, so far, is compatible with the properties of the SM Higgs. However, the SM is not the most general low-energy dynamics for the minimal electroweak symmetry breaking sector with three Goldstone bosons and one light scalar. By using non-linear effective Lagrangian for these four particles we study different processes at one-loop precision, identifying the counterterms needed to cancel the divergences. Then we apply the IAM unitarization method on the partial waves, both to make more realistic  predictions which could be tested at the LHC  and to discuss the limitations of the one-loop computations. The studied processes are the elastic scattering amplitude for both the longitudinal components of the gauge bosons $V=W,\,Z$ and the $h h \rightarrow h h$, as well as the inelastic $VV\rightarrow h h$. 
\end{abstract}


\end{frontmatter}

Two years ago  ATLAS~\cite{ATLAS} and CMS~\cite{CMS}  found a light Higgs-like boson which in principle could  make the Standard Model (SM) complete. In that case we would have a weakly interacting Electroweak Symmetry Breaking Sector (EWSBS) with three would-be Goldstone bosons $\omega^a$ (related to the longitudinal modes of vector bosons $W^{\pm}$ and $Z$) and a light scalar (the Higgs boson).
However it  is interesting to realize that so far there is a mass gap~\cite{searches} until $600$--$700$\,GeV for Higgs-like particles, and even higher for additional vector bosons. Such gap naturally suggest that the found Higgs-like boson could be understood as an additional Goldstone boson (composite state) resulting from a strongly interacting EWSBS dynamics. For example  from a model like the MCHM (Minimal Composite Higgs Model) based on the SO(5)/SO(4) coset, or from the spontaneous breaking of scale invariance symmetry (dilaton models).
In recent works~\cite{OURS}, the authors used a non-linear  effective Lagrangian with three would-be Goldsone bosons $\omega^a$ ($a=1,2,3$) and a Higgs-like light scalar $h$, given by:
\begin{eqnarray} \label{Lagrangian}
{\cal L} & = & \left( 1\! +\! 2 a \frac{h}{v} +b \frac{h^2}{v^2}\right)
\frac{\partial_\mu \omega^a \partial^\mu \omega^b}{2}
\left( \delta^{ab}\!+\! \frac{\omega^a\omega^b}{v^2} \right)   
\nonumber  \\ 
 & + &  \frac{4 a_4}{v^4}\left( \partial_\mu \omega^a\partial_\nu \omega^a \right)^2 
+ \frac{4 a_5}{v^4}\left( \partial_\mu \omega^a \partial^\mu \omega^a\right)^2  
 \nonumber  \\
 & + & \frac{2 d}{v^4} \partial_\mu  h \partial^\mu h\partial_\nu \omega^a  \partial^\nu\omega^a
+\frac{2 e}{v^4}\left( \partial_\mu h \partial^\mu \omega^a\right)^2
  \nonumber\\
 & + & \frac{1}{2}\partial_\mu h \partial^\mu h +\frac{g}{v^4} (\partial_\mu h \partial^\mu h)^2  
\end{eqnarray}

This Lagrangian can reproduce the low-energy dynamics of many models for different values of the parameters. Particular cases are the SM, $a^2=b=1$; the now experimentally excluded Higgsless Electroweak Chiral Lagrangian~\cite{Appelquist:1980vg,Longhitano:1980iz}, $a^2=b=0$; the SO(5)/SO(4) MCHM~\cite{ref_MCHM}, $a^2=1-(v/f)^2$, $b=1-2(v/f)^2$; and the simplest dilation models~\cite{ref_Dilaton}, $a^2=b=1-(v/f)^2$. Here, the $f$ parameter is a new energy scale whose precise meaning depends on the model considered.


Some experimental constraints on the different parameters can be found for example in~\cite{const_Giardino}. As a state with two Higgses $h$ has not yet been detected, there is little to say about the $b$ parameter. Thus the strongest experimental constraints are set on the $a$ parameter. At $2\sigma $  confident level, $a\in (0.70,\,1.1)$ (CMS, \cite{const_CMS}), or $a\in (0.87,\,1.3)$ (ATLAS, \cite{const_ATLAS}).

As the regime of energies where new physics is allowed starts at $600$--$700$\,GeV, we have dismissed the masses of vector bosons $\omega^a$ and scalar $h$, and used the Equivalence Theorem~\cite{ET}, relating the Goldstone bosons $\omega^a$ and the longitudinal components of the gauge bosons $W^a_L$:
\begin{equation}
T(\omega^a\omega^b \rightarrow \omega^c \omega^d) = T(W_L^a W_L^b\rightarrow W_L^c W_L^d)
+ O\left(\frac{M_W}{\sqrt{s}}\right)
\end{equation}
On the other hand our effective theory is not supposed to work for energies beyond $ 4 \pi v$. Therefore the range of applicability of our results is  $M_W, M_h\ll E\ll 4\pi v\simeq 3\,{\rm TeV}$.

From the the  Lagrangian above, and using standard techniques, it is possible to compute finite one-loop amplitudes for the processes considered here in terms of the parameters $a$ and $b$ and the renormalized couplings $a_4, a_5, d, e$ and $g$   (see ~\cite{OURS}). However it is well known that the one-loop amplitudes only fulfill unitarity perturbatively. For this reason we use dispersion relations to improve this unitarity behaviour. This can lead to different unitarization methods. In ref.~\cite{OURS} we have checked some of these methods at tree level.
We found that even while they produce slightly different numerical results, qualitatively most of them agree. At the one-loop level the so called Inverse Amplitude Method (IAM,~\cite{IAM}) seems to be the most appropriate. By using it we can reorganize the one-loop results so that we get:
\begin{equation}
\Imag\tilde{t}_\omega = \tilde{t}_\omega\tilde{t}^*_\omega + \tilde{t}_{\omega h}\tilde{t}^*_{\omega h},
\end{equation}
where $\tilde{t}_\omega$ and $\tilde{t}_{\omega h}$ are, respectively, the (iso)scalar partial waves of the exact reaction matrices associated with the processes $\omega\omega\rightarrow\omega\omega$ and $\omega\omega\rightarrow hh$.

\begin{figure}[!h]
\null\hfill\includegraphics[width=.33\textwidth]{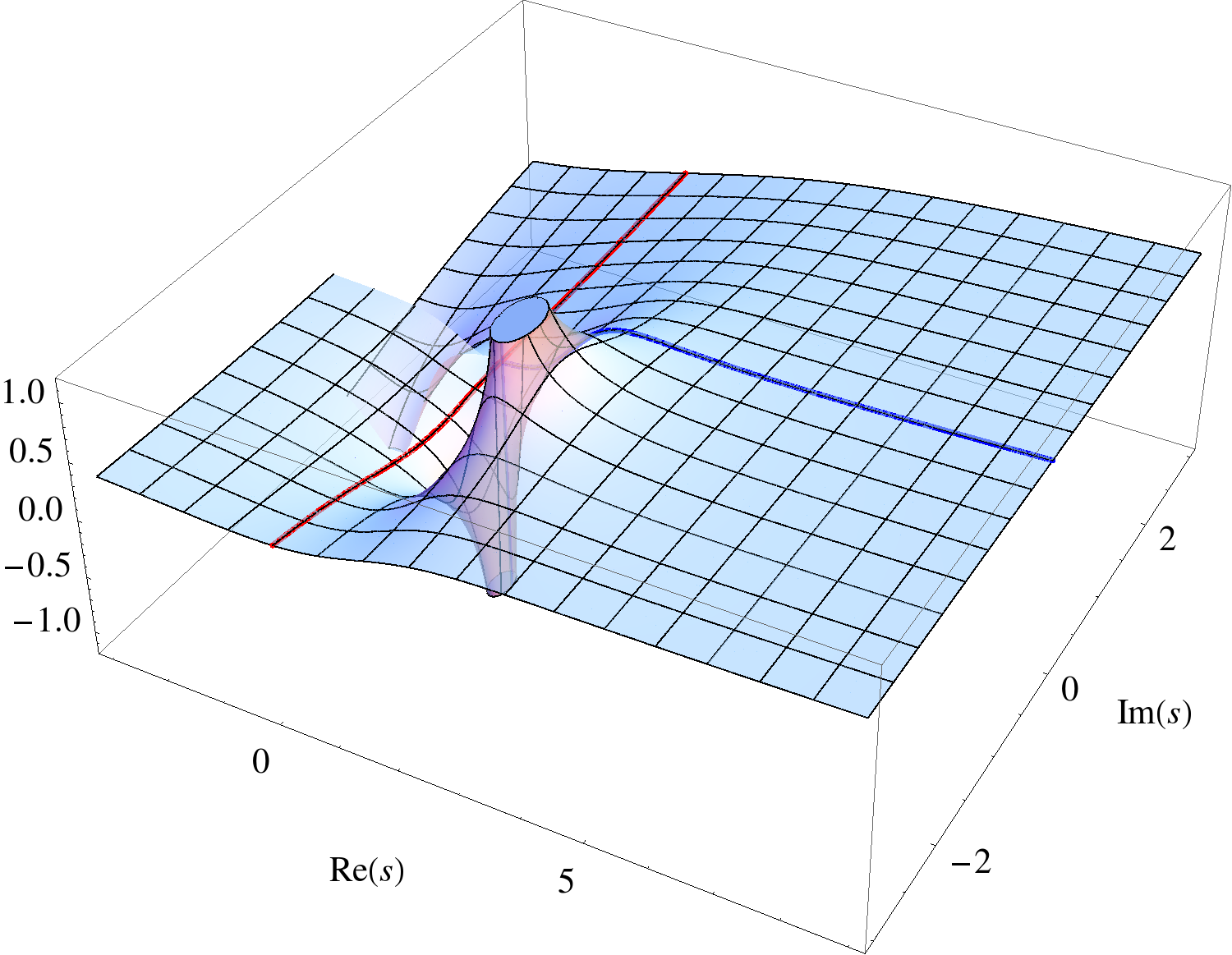}\hfill\null\\
\null\hfill\includegraphics[width=.33\textwidth]{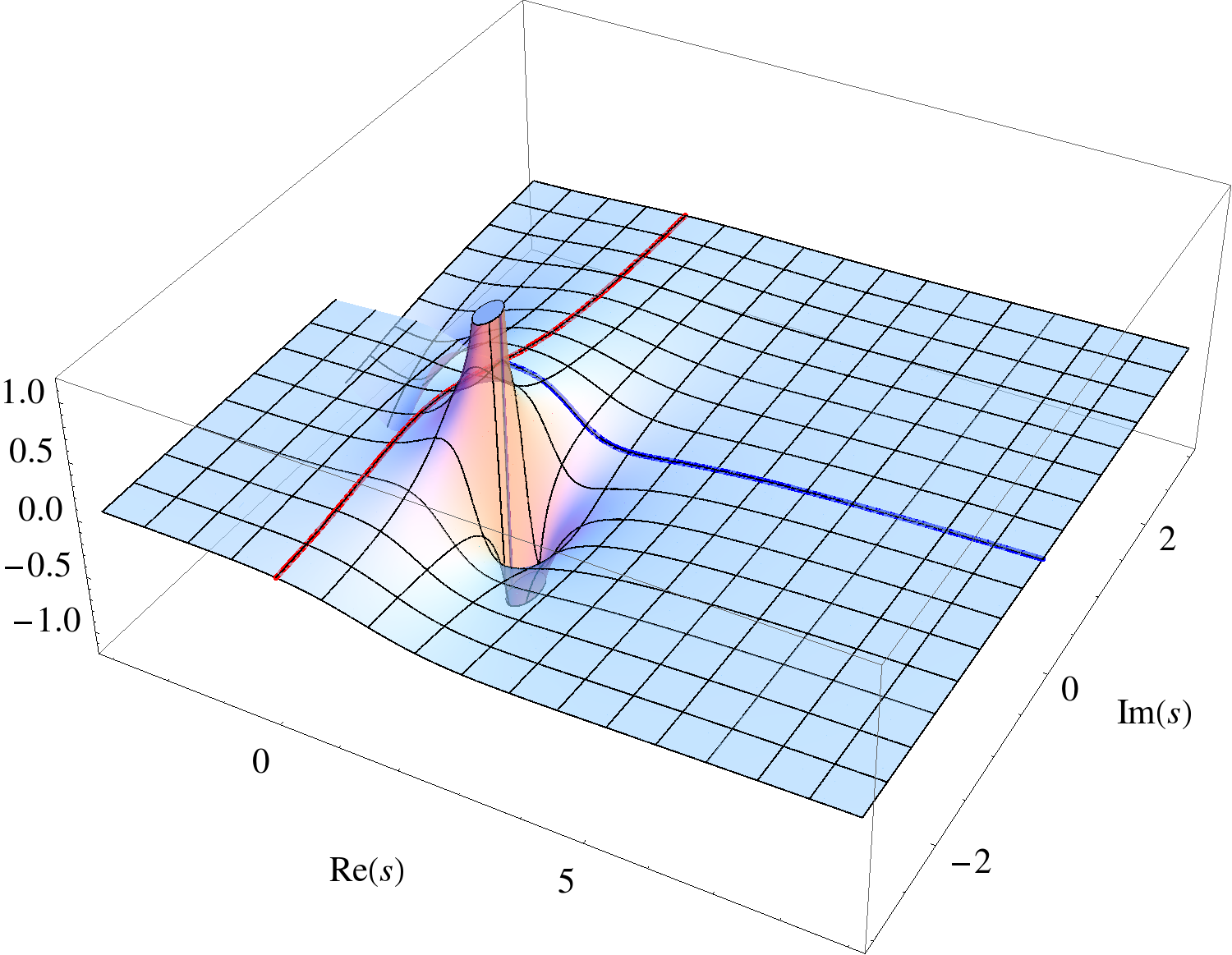}\hfill\null
\caption{Imaginary part of the unitarized partial waves in the second Riemann sheet in terms of the Mandelstam variable $s$ for $a=1$, $b = 2$.
From top to bottom, the elastic $A^{\rm IAM}$ and inelastic $M^{\rm IAM}$ amplitudes.
These amplitudes are different but still they show a pole at the same point of the second
Riemann sheet that could be understood as a new resonance.%
}\label{cross_chan}
\end{figure}

Proper unitarization methods as the IAM not only produce unitary amplitudes but also can give rise to poles in second Riemann sheet which can be understood as dynamical resonances under some circumstances. Thus the mass and width of these resonances depend on the values of the parameters and couplings of the model.
In Ref.~\cite{espriu:2013}  the authors  considered the case of varying $a_4$ and $a_5$ and how this influences the resonances appearing in the $W_LW_L$ elastic scattering in different channels. Here we concentrate on the effect of the interference with the inelastic $\omega\omega\rightarrow hh$ (for $a^2=b\neq 1$). As it can be seen in fig.~\ref{cross_chan}, this mixing could generate a resonance in the $\omega\omega\rightarrow\omega\omega$ channel even for $a=1$ i.e. when the elastic channel is suppressed 
at the leading order ~\cite{OURS,Khemchandani:2011et}.

\begin{figure}[!h]
\null\hfill\includegraphics[width=.4\textwidth]{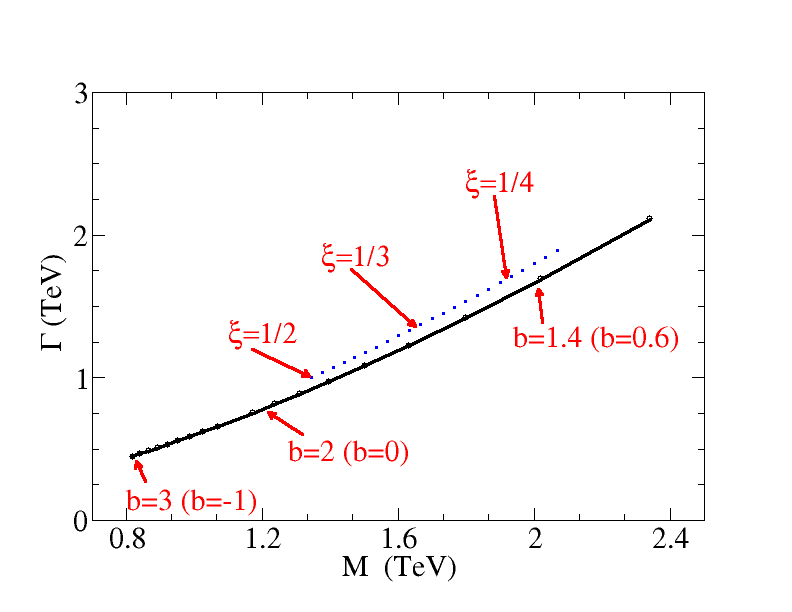}\hfill\null
\caption{Resonance mass and width, for different values of $b$ with $a^2=1$ fixed (lower curve) and for the MCHM model $a=\sqrt{1-\xi}$ and $b=1-2\xi$, $\xi$ being $v^2/f^2$ (upper dotted curve, blue online).%
}\label{pole_position}
\end{figure}

In fig.~\ref{pole_position} we show the properties of this resonance  for different values  of the $a$ and $b$ parameters. The two particular models considered there are $a=1$ (that is, no direct $\omega\omega\rightarrow\omega\omega$ strong interaction at the leading order) and the MCHM ($a=\sqrt{1-\xi}$, $b=1-2\xi$ and $\xi=v^2/f^2$).

\begin{figure}[!h]
\null\hfill\hspace{-.5cm}\includegraphics[width=.52\textwidth]{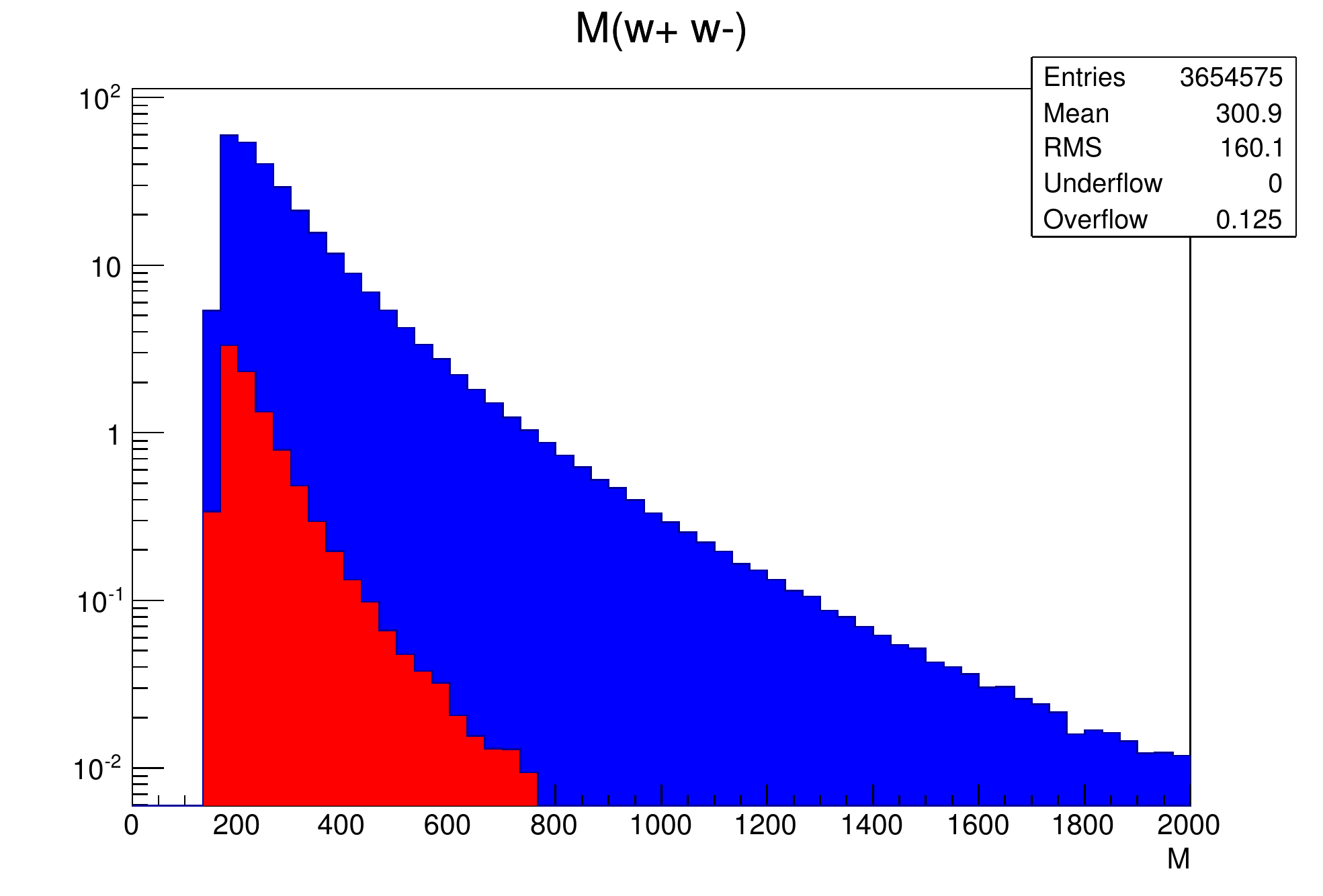}\hfill\null\vspace{-.8cm}
\caption{Production of vector bosons $W^+W^-$ (blue) vs. longitudinal modes $W_L^+W_L^-$ (red) in the SM, with $\sqrt{s}=13\,{\rm TeV}$, $L=10\,{\rm fb}^{-1}$). A strongly interacting model as those exposed here would predict an increase of the production of $W_L^+W_L^-$ for $\sqrt{s}\sim 1\,{\rm TeV}$%
}\label{MC}
\end{figure}

The EWSBS has at least a light Higgs-like boson, and three would-be Goldstone bosons (related to the longitudinal modes of $W^\pm$ and $Z$).  The case for a strongly interacting EWSBS, suggested by the gap observed at the LHC, allows for a description of the low-energy dynamics of this sector based in a non-linear effective Lagrangian similar to those used in ordinary ChPT for low-energy hadron dynamics.
Different scenarios  for the electroweak symmetry breaking can be modelized by tuning the parameters appearing in the Lagrangian.  For most of the parameter space (including models like dilaton, composite Higgs,..), we would have a strongly interacting EWSBS, giving rise typically to dynamical resonances as it happens also in hadron physics~\cite{Khemchandani:2011et}. In order to deal with such strongly interacting situations for the EWSBS, one needs to use  some of the available unitarization methods, which were developed for hadronic physics a long time ago. 

From the experimental point of view the most clear signature of a strongly interacting EWSBS would be an increment on the longitudinal W and Z bosons production in the future LHC runs. In fig.~\ref{MC}, we have simulated the production of $W^+ W^-$ vector bosons, both transverse and longitudinal for  the particular case of the SM ($a=b=1$ and the rest of the parameters vanishing). This is the only case in which we have weak interactions. Any variation from these values of the parameters produces strong interactions.
As it can be seen in fig.~\ref{MC}  the production of longitudinal  $W^+_L W^-_L$ at high energies is very small in the SM. In other models we expect a strong production. Therefore the detection of any $W$ and $Z$ pairs excess over the expected SM signal would be a signal of a strongly interacting EWSBS
and could even provide some information about the parameters of the effective theory. Of course it is still too soon to have a detailed picture of how this could be but work in that direction is in progress.

RLD thanks the hospitality of the NEXT institute and the high energy group at the University of Southampton. The authors thank J. R. Pel\'aez for bringing ref.~\cite{Khemchandani:2011et} to their attention. We acknowledge the computer resources, technical expertise, and assistance provided by the BCS and the Tirant supercomputer staff at Valencia. Work supported by spanish grants FPA2011-27853-C02-01 and BES-2012-056054 (RLD).




\begin{thebibliography}{00}

\bibitem{ATLAS}  G. Aad et al. (ATLAS Collaboration),
Phys. Lett. B{\bf 716}, 1; Report No. ATLAS-CONF-2012-168 (2012).  
\bibitem{CMS} S. Chatrchyan et al. (CMS Collaboration),
Phys. Lett. B{\bf 716}, 30 (2012); Report No. CMS-HIG-12-015.

\bibitem{searches}  
  G.~Aad {\it et al.}  [ATLAS Collaboration],
  Phys.\ Lett.\ B {\bf 712}, 22 (2012),
  Phys.\ Lett.\ B {\bf 722}, 305 (2013);
  S.~Chatrchyan {\it et al.}  [CMS Collaboration],
  Phys.\ Lett.\ B {\bf 704}, 123 (2011)


\bibitem{OURS}
Rafael L. Delgado, Antonio Dobado, Felipe J. Llanes-Estrada, (2014),
  arXiv:1408.1193 [hep-ph]; JHEP {\bf 1402} (2014) 121; J.Phys. G {\bf 41} (2014) 025002


\bibitem{Appelquist:1980vg}
  T.~Appelquist and C.~W.~Bernard,
  Phys.\ Rev.\ D {\bf 22} (1980) 200.

\bibitem{Longhitano:1980iz}
  A.~C.~Longhitano,
  Phys.\ Rev.\ D {\bf 22} (1980) 1166;

\bibitem{ref_MCHM}
K. Agashe, R. Contino and A. Pomarol, Nucl. Phys. B {\bf 719}, 165 (2005);
R. Contino, L. Da Rold and A. Pomarol, Phys. Rev. D {\bf   75}, 055014 (2007);
R. Contino,  D.  Marzocca,  D. Pappadopulo and R. Rattazzi,
JHEP {\bf 1110} (2011) 081;
D. Barducci et al. JHEP {\bf 1309}, 047 (2013).

\bibitem{ref_Dilaton}
    E. Halyo,
    Mod. Phys. Lett. A {\bf 8} (1993) 275;
%
    W. D. Goldberger, B. Grinstein and W. Skiba,
    Phys. Rev. Lett. {\bf 100} (2008) 111802.


\bibitem{HerreroCillero}
R.L. Delgado, A. Dobado, M.J. Herrero, J.J. Sanz-Cillero, JHEP {\bf 1407} (2014) 149

\bibitem{const_Giardino}
   Giardino, P.P., Aspects of LHC phenomenology, PhD Thesis (2013), Universit\`a di Pisa
\bibitem{const_CMS}
   [CMS Collaboration], Collaboration report CMS-PAS-HIG-12-045.
\bibitem{const_ATLAS}
   G. Aad et al. [ATLAS Collaboration], Phys. Lett. B {\bf 726}, 88 (2013).

\bibitem{ET} 
J.M. Cornwall, D.N. Levin and G. Tiktopoulos, Phys. Rev. D{\bf 10} (1974) 1145;
C.E. Vayonakis, Lett. Nuovo Cim.{\bf 17} (1976) 383;
B.W. Lee, C. Quigg and H. Thacker, Phys. Rev. D{\bf 16} (1977) 1519;
M.S. Chanowitz and M.K. Gaillard, Nucl. Phys. {\bf 261} (1985) 379;
M. S. Chanowitz, M. Golden and H. Georgi, Phys. Rev. D{\bf 36} (1987) 1490;
A. Dobado J. R. Pel\'aez Nucl. Phys. B{\bf 425} (1994) 110; 
Phys. Lett.B329 (1994) 469 (Addendum, ibid, B{\bf 335} (1994) 554).


\bibitem{IAM} 
  A.~Dobado, M.~J.~Herrero and T.~N.~Truong,
  Phys.\ Lett.\ B {\bf 235}, 129 (1990).
A.~Dobado and J.~R.~Pelaez,
  Phys.\ Rev.\ D {\bf 56}, 3057 (1997).

\bibitem{espriu:2013}
D. Espriu, F. Mescia and B. Yencho, Phys. Rev. D {\bf 88}, 055002 (2013);
D. Espriu and B. Yencho, Phys.Rev. D {\bf 87}, 055017 (2013).


\bibitem{Khemchandani:2011et} 
A similar effect has been suggested to occur in the $I=1/2$ resonance oscillating between $\phi N$ and $K^*\Lambda$ around 2 GeV 
  K.~P.~Khemchandani {\it et al.},
  Phys.\ Rev.\ D {\bf 83}, 114041 (2011); see also
  E.~Oset and A.~Ramos,
  Eur.\ Phys.\ J.\ A {\bf 44}, 445 (2010).

\end{thebibliography}




\end{document}